\documentclass[twocolumn,a4paper,showpacs,preprintnumbers,amsmath,amssymb,floatfix]{revtex4-1}
\usepackage{amsmath}
\pdfoutput=1
\usepackage{graphicx}
\usepackage{dcolumn}
\usepackage{bm}
\usepackage{times}
\usepackage{multirow}
\usepackage{longtable}
\usepackage{tabu}
\begin{document}

\title{Graphene quantum dots probed by scanning tunneling spectroscopy and transport spectroscopy after local anodic oxidation}


\author{%
  Markus Morgenstern\textsuperscript{\textsf{\bfseries 1}},
  Nils Freitag\textsuperscript{\textsf{\bfseries 1}},
  Aviral Vaid\textsuperscript{\textsf{\bfseries 1,2}},
  Marco Pratzer\textsuperscript{\textsf{\bfseries 1}}, and
  Marcus Liebmann\textsuperscript{\textsf{\bfseries 1}}
  }



\affiliation{%
  \textsuperscript{1}\,II. Institute of Physics B and JARA-FIT, RWTH Aachen, 52074 Aachen, Germany\\
  \textsuperscript{2}\, Department of Materials Science, Friedrich-Alexander-University Erlangen-N\"urnberg,
Martensstrasse 5-7, 91058 Erlangen, Germany\\
}

\received{XXXX, revised XXXX, accepted XXXX} 
\published{XXXX} 


%
%
%
\begin{abstract}
 Graphene quantum dots are considered as promising alternatives to quantum dots in III-V semiconductors, e.g., for the use as spin qubits due to their
 consistency made of light atoms including spin-free nuclei which both imply relatively long spin decoherene times. However, this potential has not been realized in experiments so far, most likely, due to a missing control of the edge configurations of the quantum dots. Thus, a more fundamental investigation of Graphene quantum dots appears to be necessary including a full control of the wave function properties most favorably during transport spectroscopy measurements.\\
 Here, we review the recent success in mapping wave functions of graphene quantum dots supported by metals, in particular Ir(111), and show how the goal of probing such wave functions on insulating supports during transport spectroscopy might be achieved.
\end{abstract}  
\date{\today}
%
%

\maketitle   


Graphene has moved in short time from first preparation as a small flake \cite{Geim1} towards  applications, partly already realized and partly anticipated for the near future, such as high frequency transistors \cite{Avouris}, supercapacitors \cite{supercapacitor}, or touch screens \cite{samsung} as recently reviewed by Ferrari {\it et al.} \cite{Ferrari}.
Most of the applications are based on the exceptional properties of the material including true two-dimensionality, inertness, high room temperature mobility, large thermal conductivity, and extreme mechanical breaking strength \cite{Geim}.\\
Another exciting plan is to use graphene quantum dots (QDs) as
spin qubits \cite{Trauzettel}. The basic requirement is a very long spin coherence time \cite{diVinc},
which might exist in graphene \cite{Burkard} due to the absence of hyperfine
coupling in isotopically pure material and the small spin-orbit
coupling \cite{Fabian}. First graphene QDs have been produced early on and probed by
transport measurements \cite{Ponamerenko,Stampfer,Guttinger}, meanwhile being able to determine charge relaxation times of about 60-100 ns \cite{Volk}.
However, since graphene provides no natural gap, it is difficult to control the
electron number \cite{Libisch12}.
Moreover, the 2D sublattice symmetry makes the QD properties very susceptible to the atomic edge configuration \cite{Trauzettel} in contrast to the behavior in conventional QDs made, e.g., out of GaAs. As a result, chaotic Dirac billiards have been predicted for graphene QDs \cite{ber87}
and were even claimed to be realized \cite{Ponamerenko,Wurm}, i.e. the wave functions are probably rather disordered.
To get more control on graphene QDs, the QD edges must be well defined and a more fundamental understanding of the QD properties is mandatory.
More recent transport investigations on graphene nanoconstrictions indeed support the idea of a decisive role of the edges for the
localization properties within graphene nanostructures \cite{Bischof}.\\
A direct insight into QD properties is provided by scanning tunneling spectroscopy (STS). STS can map
the squared wave functions of QDs \cite{Berndt2} and, at the same
time, determines the shape of the QD atom by atom. STS has also been applied successfully to graphene samples \cite{Morgenstern2}. However, conventional STS is restricted to conducting surfaces,
such that first investigations of QD wave functions have been performed on graphene nanoislands prepared in Ultra-High-Vacuum (UHV) on metals \cite{Subramaniam,Hamaleinen,Park,Altenburg,Leicht,Craes,Jolie}.
More peculiar quantum dots have also been probed by scanning tunneling microscopy (STM) in UHV such as the ones induced by tip-induced strain on suspended graphene areas \cite{Klimov},
the ones being present within localized areas of a quantum Hall sample \cite{Jung}, or the ones confined in the suspended areas of the strongly buckled graphene on Ru(0001) \cite{Zhang}.\\
Here, we firstly review the results of wave function mapping of graphene QDs on Ir(111),
which includes the probing of the zig-zag edges, appearing to be the energetically favorable edge type
for this substrate \cite{Lacovig}. It is found that the graphene QDs exhibit squared wave functions \cite{Subramaniam,Hamaleinen,Altenburg,Park}, which are a mixture of graphene states from the Dirac cone located in a projected band gap of the Ir(111) \cite{Petikosic} and hole-type Ir(111) surface states located around $\Gamma$ and exhibiting a strong Rashba-type spin splitting \cite{Rader}. The strength of the two contributions within the QD can be disentangled by Fourier transformation of the real space data,
if atomic resolution is achieved \cite{Leicht}. The contribution from the graphene Dirac cone state can, moreover, be enhanced by oxygen intercalation between the graphene nano-islands and the Ir(111) surface \cite{Jolie}.\\
Interestingly, the edge configuration of the graphene quantum dots on Ir(111) is quite unique by the hybridization of the graphene p$_z$ orbitals with Ir d$_{z^2}$ orbitals, the latter being located rather exactly at the Fermi level $E_{\rm F}$. This, on the one hand side, suppresses the appearance of the famous edge state at zig-zag edges \cite{Li2} and, on the other hand side, leads to a soft confinement of the graphene quantum dot states \cite{Subramaniam}. Thus, the confinement is not atomically abrupt, as one might anticipate from the abrupt end of carbon atoms at the rim of the island, but proceeds over 4-5 lattice constants.  As a result, the interference effects, coming from the coupling of the two valleys K and K' and typically making the quantum dot wave functions chaotic, are suppressed \cite{Wurm,Geim}, which, in turn, makes the wave functions rather regular.
Indeed, calculations of the atomically identical quantum dots, which are freely suspended and, thus, not prone to the interaction with the substrate, show strongly irregular wave functions distinct from the experiment. Only the inclusion of the soft confinement potential could recover the experimental results.\\
Thus, it is experimentally demonstrated that soft confinement leads to a significant regularization of states in graphene quantum dots and, thus, should be part of a strategy
to exploit the potential of graphene quantum dots, e.g., with respect to spin qubits \cite{Trauzettel}.\\
In addition, the soft confinement is able to suppress the graphene edge states.  Such states with almost flat bands at the Dirac point energy $E_{\rm D}$ are an intrinsic property of graphene zigzag edges, at least, if unsupported and unreconstructed \cite{Fujita2,Nakada,Pisani,Wassmann}. Experimental evidence for such edge states has indeed been found on different substrates such as  HOPG \cite{Kobayashi,Ziatdinov}, Si(100) \cite{Ritter},  Au(111) \cite{Tao,Magda}, Cu(001), if the graphene is laterally interfaced with BN, \cite{Park2}, arguably Pt(111) \cite{Merino}, SiC(0001) \cite{Baringhaus}, and SiO$_2$ \cite{Pan,Chae}. Spectroscopic features at the step edges of Ir(111) covered with graphene have also been interpreted in terms of graphene edge states \cite{Klusek}, albeit it is known that the graphene layer is not interrupted at the Ir step edges \cite{Coraux}.\\
Density functional theory calculations indeed show that the edge state is present on Cu(111), Ag(111), and Au(111) \cite{Li3}, but not on Ir(111) \cite{Li2,Phark2}. It is, moreover, shown that
H-terminated zig-zag edges on Au(111) are ferromagnetic, i.e. the edge state is located at $E_{\rm F}$ \cite{Li3}. Indirect evidence for such ferromagnetism of the edge state has been published \cite{Tao,Magda}, but without chemical control of the edge termination so far, $-$ being possible in principle \cite{Zhang2}. Such edge state magnetism might be exploited for a multitude of spintronic applications~\cite{Son,Kim,Wakabayashi,Wimmer}, but, on the other hand, it is detrimental for quantum dot functionality. Thus, the edge state suppression by soft confinement, as
on Ir(111), again improves the reliability of the quantum dot wave functions.\\
Interestingly, the edge state can be recovered by laterally interfacing the graphene zig-zag edges on Ir(111) with BN and subsequently intercalating Au between the graphene/BN layer and the Ir(111) surface \cite{Drost}.\\
\begin{figure}[htb]%
\includegraphics*[width=0.5\textwidth]{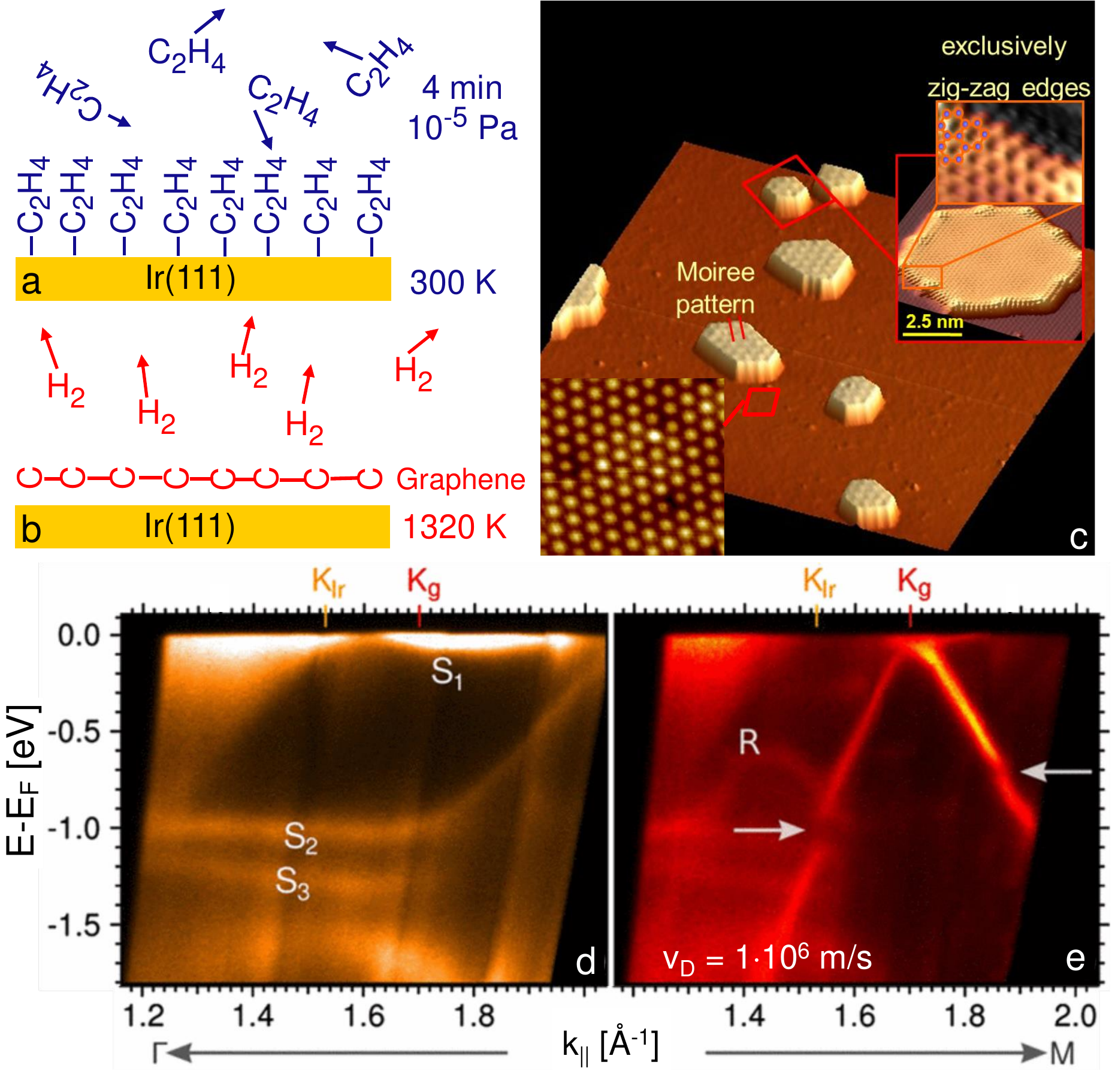}
\caption{(a) Sketch of exposure of C$_2$H$_4$ to Ir(111) at 300 K; (b) sketch of subsequent heating in order to produce graphene;
(c) 3D representation of a ($100\times100$)\,nm$^2$ STM image of Ir(111) covered by monolayer graphene islands; $V=-0.3$~V, $I=0.3$\,nA; upper inset: atomically resolved STM image of graphene island with zoom into zigzag edge having a sketch of the graphene lattice overlaid, $V=0.7$\, V, $I=20$\,nA; lower inset: zoom into Ir(111) with atomic resolution, $V=0.9$\, V, $I=0.5$\,nA; (d) ARPES spectrum of clean Ir(111) recorded along $\Gamma$M direction with positions of the K points of iridium and graphene marked as K$_{\rm Ir}$ and K$_{\rm g}$, respectively; S$_1$ $-$ S$_3$ are surface states; (e) ARPES spectrum of Ir(111) covered by graphene along the same direction as in (a); Dirac velocity $v_{\rm D}$, Dirac cone replica (R) and minigaps at the crossing points of original and replicated Dirac cones (arrows) are marked. \cite{Subramaniam,Petikosic}
((c), (d) courtesy of C. Busse, University of Cologne)}
\label{Fig1}
\end{figure}
Albeit these STS studies of graphene QDs on metals open the door towards a better understanding of QD wave function properties, they are still relatively remote from the
transport investigations \cite{Ponamerenko,Stampfer,Bischof}. A direct imaging of the wave functions responsible for the transport signatures would be more favorable. This goal is also pursued by scanning gate microscopy \cite{Schnez}, but the achieved spatial resolution so far only allows to locate the center of the corresponding wave functions without any details \cite{Schnez,Conolly}. Combining transport and scanning tunneling microscopy is possible in principle, if the nanostructures are produced without resist as, e.g., by anodic oxidation using an atomic force microscope (AFM) \cite{Campbell}. This technique has been shown to work on graphene, too \cite{Weng}, and to produce operating quantum dot
structures \cite{Neubeck}. Besides cuts, where the graphene is removed \cite{Weng}, the graphene can also be oxidized \cite{Masubushi2} or hydrogenated \cite{Byun}, which eventually could lead to smaller, insulating device boundaries. Importantly, this method does not leave any resist on the surface, which would disturb the STS measurements significantly due to a remote charging of the resist by the tip, which electrostatically would back-act on the area probed by STS \cite{Geringer2}.
Combining this AFM based structuring method
with a contacting method, which does not use any resist either as, e.g., microsoldering by In \cite{Girit,Geringer2}, provides an adequate nanostructure to be probed both, by
STS and transport spectroscopy simultaneously.\\
In order to scan across the insulting barriers, the scanning tip has to be mounted on a tuning fork \cite{Albers} , such that the feedback can operate in scanning force mode while tunneling, or, alternatively, a capacitive feedback must be used \cite{Andrei}.\\

%
\begin{figure*}[htb]%
\includegraphics*[width=\textwidth]{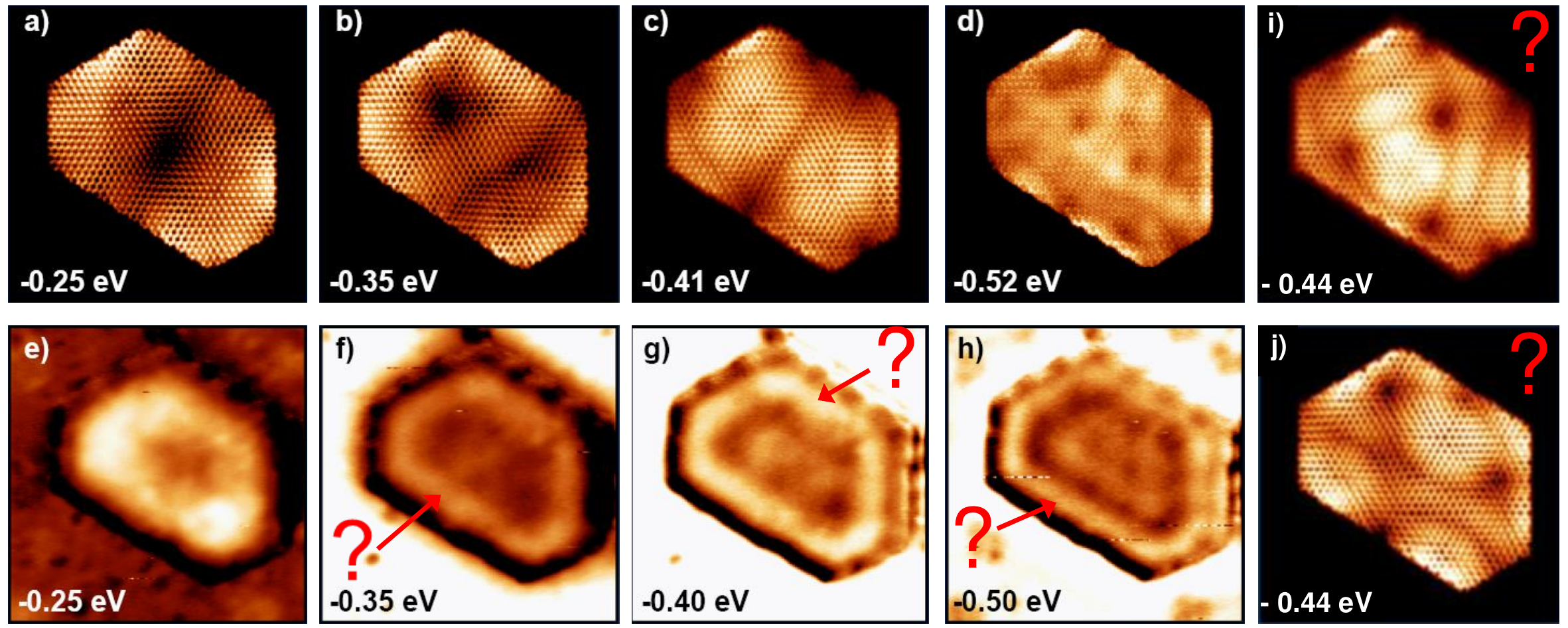}
\caption{(a)-(c) Calculated squared wave functions of the island probed in (e)-(h) using third-nearest neighbor tight-binding calculations which ignore the substrate; energies are marked;  (d) calculated LDOS of the same island (consisting of 6 wave functions) at the energy marked; (e)-(h) $dI/dV$ maps of the island at different energies $E=V\cdot e$ as marked; $I=0.3$ nA, $U_{\rm mod}=10$ mV; question marks mark the bright rim observed in experiment, but not in the TB calculations;  (i)-(j) calculated squared wave functions
of the same island, which are not found in the experiment as marked by question marks; all calculations are with abrupt edge. \cite{Subramaniam}
}
\label{Fig2}
\end{figure*}
\section{Graphene Quantum dots on Ir(111)}
\subsection{Wave function mapping}

Monolayer graphene islands acting as QDs can be prepared by exposing clean Ir(111)
for 4 min to a pressure of 10$^{-5}$ Pa of C$_2$H$_4$ at 300 K and subsequent annealing to 1320\,K (30 s) \cite{NDiyae} as sketched in Fig. \ref{Fig1}(a) and (b). The resulting islands (graphene QDs) have diameters of $2-40$\,nm and are completely enclosed by zig-zag edges as shown in Fig. \ref{Fig1}(c).\\
Fortunately, Ir(111) exhibits a projected band gap in the region of the graphene Dirac cone \cite{Petikosic} such that graphene QD states can be observed \cite{Subramaniam,Hamaleinen,Park}. The band gap located between the surface states S$_1$ and S$_2$ is visible in the angular resolved photoelectron spectroscopy (ARPES) data shown in Fig. \ref{Fig1}(d). After preparing a complete monolayer of graphene, the Dirac cone of graphene becomes visible within that band gap showing the well-known linear dispersion with a Fermi velocity of $v_{\rm D}=(1.0\pm0.05)\cdot10^6$ m/s \cite{Petikosic,Jolie} (Fig. \ref{Fig1}(e)). The moir{\'e} superstructure between graphene and Ir(111) induces replica of the Dirac cone (marked by R) and corresponding minigaps at their crossing points (arrows) located at higher binding energy \cite{Petikosic}. More importantly, the extrapolated Dirac point is at $0.1$ eV above $E_F$ \cite{Petikosic}, overlapping with the Ir surface state S$_1$, which has d$_{z^2}$ character \cite{Li2}, thus, the S$_1$ orbitals point directly to the p$_z$ orbitals of graphene. Consequently, hybridization of S$_1$ with the Dirac cone and, thus, a gapping of the Dirac cone at E$_{\rm D}$ is likely and has indeed been found in density functional theory (DFT) calculations \cite{Subramaniam,Li2}. Moreover, the projected band gap does only exist for the hole part of the band structure, while the electron part overlaps with bulk bands from Ir(111) \cite{Subramaniam}.\\

The local density of states (LDOS) of the QDs can be mapped by STS in constant current mode \cite{Subramaniam,Hamaleinen,Park}. We use a STM operating at $T=6$ K \cite{Mashoff1} and a lock-in technique with modulation frequency $\nu=1.4$\,kHz and amplitude $V_{\rm mod}=10$\,mV resulting in an energy resolution $\delta E_{\rm STS}\approx\sqrt{(3.3\cdot k_BT)^2+(1.8\cdot eV_{\rm mod})^2}=18$\,meV \cite{Morgenstern}.
This energy resolution is much better than the natural peak width of the confined states of $\delta E = 0.1-0.4$ eV, which increases linearly with $|E-E_{\rm F}|$. For $dI/dV$ curves, we stabilize the tip at sample voltage $V_{\rm stab}$ and current $I_{\rm stab}$ before switching off the feedback.
Fig. \ref{Fig2}(e)-(h) show the LDOS patterns observed for a particular graphene QD in comparison with third-nearest neighbor tight binding (TB) calculations, which neglect any influence of the substrate (Fig. \ref{Fig2}(a)-(d)). The exact atomic configuration of the graphene QD deduced from the STM data has been used for the calculations.
While single fourfold degenerate wave functions contribute at lower binding energy (Fig. \ref{Fig2}(a)-(c)), the LDOS at higher binding energy consists of six overlapping fourfold degenerate wave functions.
A reasonable correspondence of the wave function symmetries between TB results and STS results is found, but there are two decisive differences.\\
Firstly, the experimental LDOS exhibits a bright band at the rim of the island (question marks in Fig. \ref{Fig2}(f)-(h)) which is not present in the TB data. Secondly, the strongly disordered wave functions expected from the sublattice symmetry breaking at the rim of the QD \cite{Libisch12,Nakada} is found in the TB calculations (Fig. \ref{Fig2}(i) and (j)), but not in the STS data.\\
The first discrepancy is related to the penetration of the sp-type Ir surface state into the graphene QD, while the second one is due to the soft confinement resulting from the hybridization of the graphene p$_z$ orbitals with the Ir(111) d$_{z^2}$ surface state.
\begin{figure}[htb]%
\includegraphics*[width=0.5\textwidth]{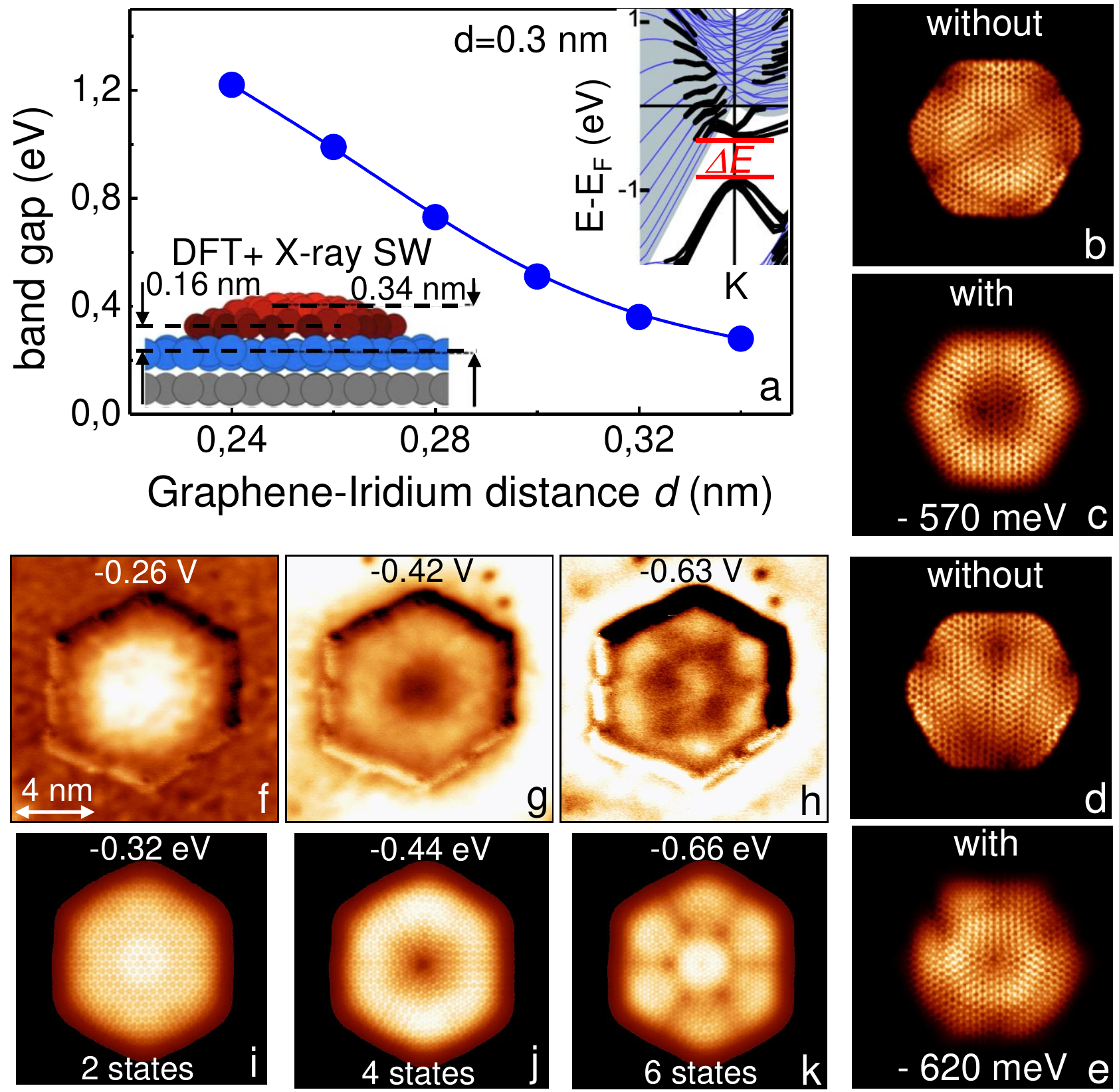}
\caption{(a) Energy gap $\Delta E$ at the Dirac point $E_{\rm D}$ versus graphene-Ir distance $d$ as deduced from DFT calculations;  upper right inset: band structure around $E_{\rm D}$ for $d=0.3$ nm  with $\Delta E$ indicated, grey area corresponding to projected bulk bands of Ir and thick black lines mark graphene states; lower left inset: sketch of a graphene island (red) on Ir(111) (blue grey) with distances marked according to DFT calculations and X-ray standing wave data \cite{Lacovig}; (b)-(e) LDOS ($=|\Psi|^2$) deduced from TB calculations for individual confined states; (b), (c) $E-E_{\rm F} = -570$ meV; (d), (e) $E-E_{\rm F} = -620$ meV; (b), (d) calculated with abrupt edges; (c), (e) calculated with additional soft edge potential; (f)-(h) $dI/dV$ images recorded at energies $E=V\cdot e$ as marked, $I=0.2$\,nA, $V_{\rm mod}=10$\,mV; (i)-(k) LDOS maps from TB calculations with soft edge potential at energies indicated, number of states contributing to the LDOS are marked. \cite{Subramaniam} ((a) courtesy of R. Mazzarello, RWTH Aachen University, (b)$-$(e) and (e)$-$(k) courtesy of F. Libisch, TU Vienna)
}
\label{Fig3}
\end{figure}

In order to take the hybridization into account, we performed density functional theory calculations of extended graphene sheets with different distance between the graphene layer and Ir(111). The resulting band structure (see inset of Fig. \ref{Fig3}(a)) exhibits a band gap $\Delta E$ at the Dirac point of the graphene related states. This $\Delta E$ increases with decreasing graphene-Iridium distance $d$ as displayed in Fig. \ref{Fig3}(a). The change of this distance at the rim of a graphene island has been deduced previously by favorable comparison of DFT calculations and x-ray standing wave experiments \cite{Lacovig}. It is located within the outer 1 nm of the graphene island and changes from 0.16 nm at the last atom to 0.34 nm in the interior of the graphene island as shown in the lower right inset of Fig. \ref{Fig3}(a).
We take this into account by an on-site potential $V_i$ at each lattice site $i$ in the TB calculation reading \cite{ber87}:
\begin{equation}
V_i=\frac{\Delta E(d)}{2}\cdot \sigma_z
\end{equation}
with $\sigma_z$ being the Pauli Matrix acting on the sublattice degree of freedom. Figure \ref{Fig3}(b)-(e) demonstrate that this on-site potential leading to soft confinement across about 5 lattice constants indeed removes the irregularities of the squared QD wave functions. Reasonable agreement between STS data and the TB data with soft confinement
is achieved for smaller islands (Fig. \ref{Fig3}(f)-(k)), where the outer rim of Fig. \ref{Fig2}(f)-(h) appears to be less strong. Notice that the moir{\'e} potential, known from the size of the minigaps (see Fig. \ref{Fig1}(e)), is included as an on-site potential, too \cite{Subramaniam}.\\
\begin{figure}[htb]%
\includegraphics*[width=0.5\textwidth]{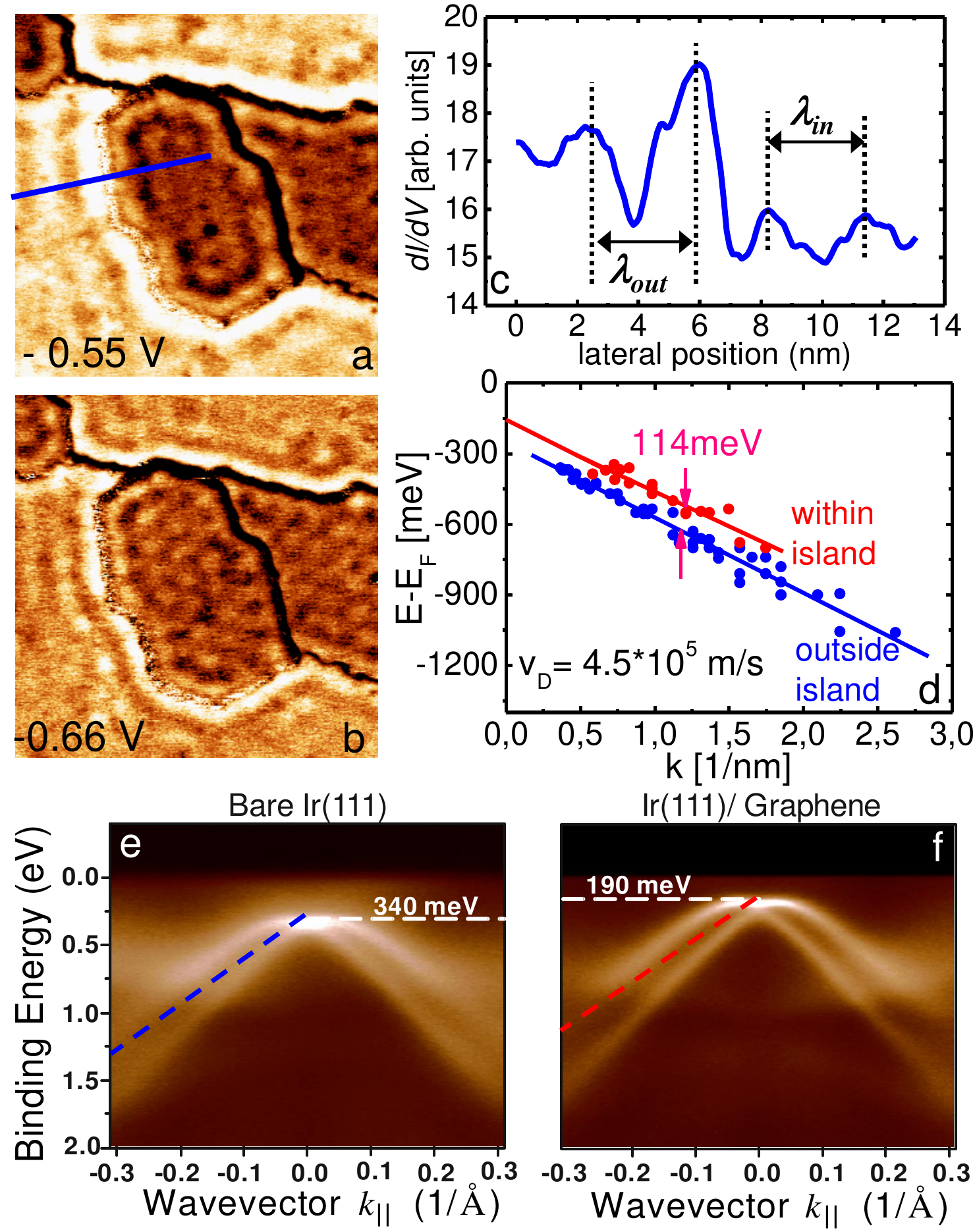}
\caption{(a)-(b) $dI/dV$ maps of a graphene QD recorded at the voltages marked; $27\times30$ nm$^2$, $I=0.5$\,nA, $V_{\rm mod}=10$\,mV; blue line marks the orientation of the profile line shown in (c); (c) profile line along the line marked in (a)
 with deduced wave lengths $\lambda_{\rm out}$ ($\lambda_{\rm in}$) outside (inside) the QD indicated; (d) resulting dispersion relations $E(\Delta k =\pi/\lambda_{\rm in/out})$ inside (red) and outside (blue) of the QDs as deduced from several QDs; full lines are linear fits with resulting $v_{\rm D}$ and relative energy offset marked; (e) ARPES data of Ir(111) showing a Rashba split hole-type surface state, energy maximum (binding energy: 340 meV) is marked; the blue fit line from (d) (dashed line) is overlaid; (f) ARPES data of Ir(111) covered with a complete monolayer of graphene; energy maximum (binding energy: 190 meV) is marked; the red fit line from (d) (dashed line) is overlaid  \cite{Rader}. \cite{Subramaniam} ((e), (f) courtesy of A. Varykhalov, Helmholtz-Zentrum, Berlin)
}
\label{Fig4}
\end{figure}
In larger islands as in Fig. \ref{Fig4}(a) and (b), the bright band at the rim of the QDs develops into a standing wave at the rim, which is also apparent with similar wave length at the outer side of the island. The wave length dependence on energy $E=e\cdot V$ is determined everywhere, where at least two maxima of the standing wave can be discriminated as shown in Fig. \ref{Fig4}(c). The wave length of a standing wave $\lambda$ is related to the wave number $k$ of the electronic state by $k=\pi/\lambda$ giving access to the dispersion relation $E(k)$ of the corresponding states inside and outside of the graphene QD. It is shown in Fig. \ref{Fig4}(d) revealing the same slope inside and outside the island. Assuming a
linear dispersion $E=\hbar\cdot v_{\rm D}\cdot k$, one reveals a Dirac velocity $v_{\rm D}=(4.5\pm 0.5)\cdot 10^5$ m/s, which is less than half of the value known for the graphene layer on Ir(111) (see Fig. \ref{Fig1}(e)). Moreover, the crossing point of the fit lines in Fig. \ref{Fig4}(d) are at $0.15-0.3$ eV below $E_{\rm F}$, while the extrapolated Dirac point of graphene is at $0.1$ eV above $E_F$. Thus, the standing waves are not related to the Dirac cone of graphene. In contrast, they fit nicely to an Ir(111) surface state of sp-type called S$_0$ and being located around $\Gamma$, which is shown in Fig. \ref{Fig4}(e) and (f) \cite{Rader}. S$_0$ shows a strong Rashba-type spin splitting and a nearly linear dispersion away from $\Gamma$. It survives the coverage of a graphene monolayer and is rigidly shifted by about 0.15 eV upwards. It is well known that the standing waves observed in STS are not susceptible to the Rashba-type spin splitting \cite{Petersen}, such that the wave length of a standing wave is related to the average $k$ of both bands. The averaged slope of both bands corresponds to $v_{\rm D}=5.5\cdot 10^5$ m/s and crossing points with $\Gamma$ at $0.1-0.25$ eV below $E_{\rm F}$ in good agreement with the standing wave data obtained by STS at the graphene QD rims. The relative energy shift of 0.15 eV upwards after graphene coverage is also in agreement with the STS data. The fit results from the STS data are overlaid as dashed lines in Fig. \ref{Fig4}(e) and (f) substantiating the good agreement. Remaining small differences are probably related to changes of the electrostatics at the rim of the islands with respect to the extended film probed by ARPES.\\
\begin{figure}[htb]%
\includegraphics*[width=0.5\textwidth]{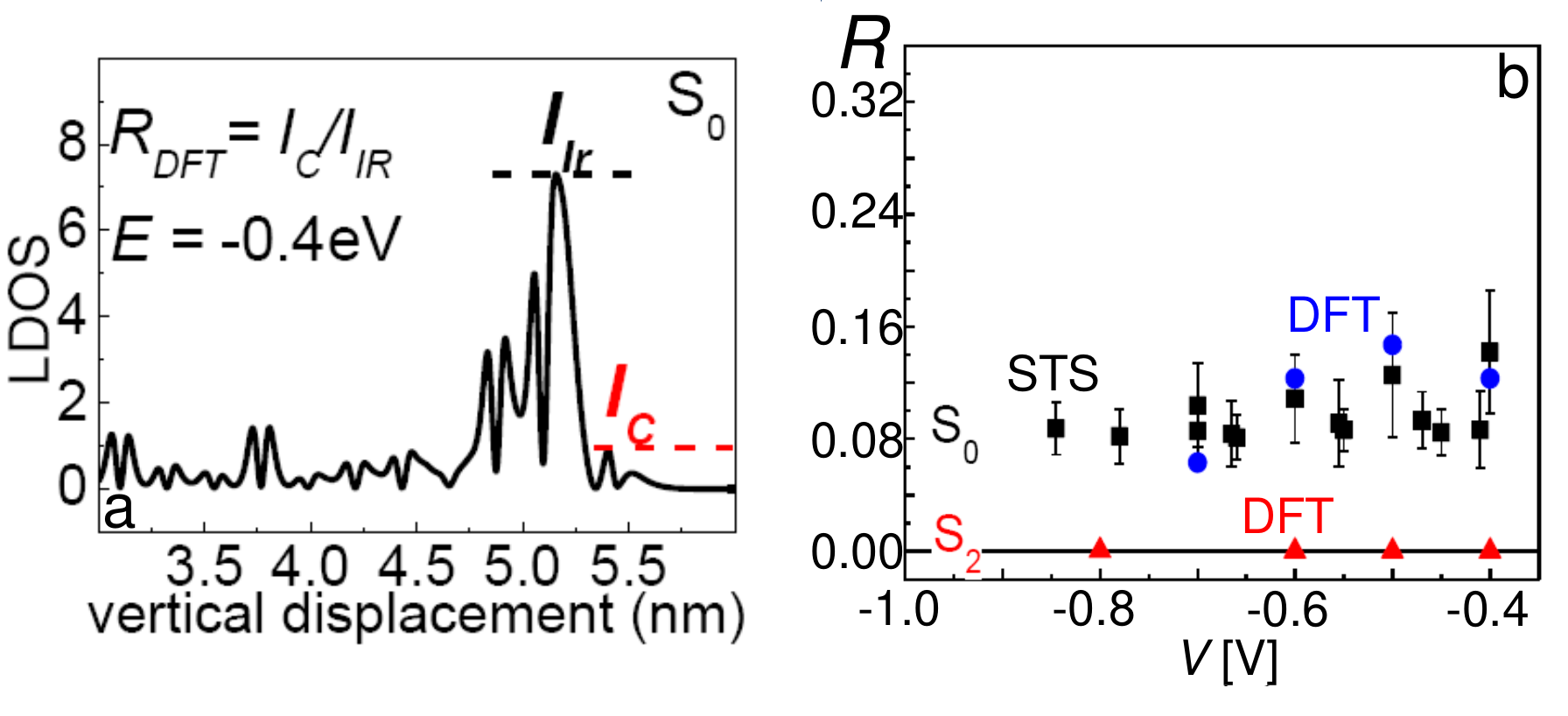}
\caption{(a) Calculated LDOS of S$_0$ for Ir(111) covered with graphene at $E=-0.4$ eV along the direction perpendicular to the surface; the intensity in the Ir surface layer $I_{\rm Ir}$ and the intensity within the graphene layer $I_{\rm C}$ as used for the determination of $R=I_{\rm C}/I_{\rm Ir}$ are marked; (b) relative intensity $R=I_{\rm C}/I_{\rm Ir}$ of S$_0$ and S$_2$ in graphene as deduced from STS data (squares) and from DFT calculations (S$_0$: circles, S$_2$: triangles). \cite{Subramaniam} (Courtesy of R. Mazzarello, RWTH Aachen University)
}
\label{Fig5}
\end{figure}
In order to corroborate the conclusion that S$_0$ contributes to the LDOS patterns of the graphene QDs, we analyzed the vertical distribution of the S$_0$ states as calculated by DFT. As shown in Fig. \ref{Fig5}(a), S$_0$ indeed penetrates into the graphene layer with a few percent of its intensity. We compared the strength of S$_0$ in the surface layer of Ir $I_{\rm Ir}$ and the strength of S$_0$ in the graphene layer $I_{\rm C}$ giving a ratio $R_{\rm DFT} = I_{\rm C}/I_{\rm Ir}$. This ratio displayed in Fig. \ref{Fig5}(b) is 10-15 \%, while it is negligible for other d-type surface states of Ir(111) as, e.g., S$_2$. The ratio has been compared with the relative amplitudes of the standing waves inside and outside the graphene QD (see Fig. \ref{Fig4}(c)) after recalibrating for different tip-surface distances above Ir(111) and above graphene being deduced from the known height of graphene above Ir(111) (0.34 nm, Fig. \ref{Fig3}(a)) and recorded $I(z)$ curves. As shown in Fig. \ref{Fig5}(b), the agreement between STS and DFT data is very good, substantiating that S$_0$ contributes to the LDOS patterns in the QD.

\subsection{Disentangling graphene Dirac cone states and Ir surface states}

\begin{figure*}[htb]%
\includegraphics*[width=0.5\textwidth]{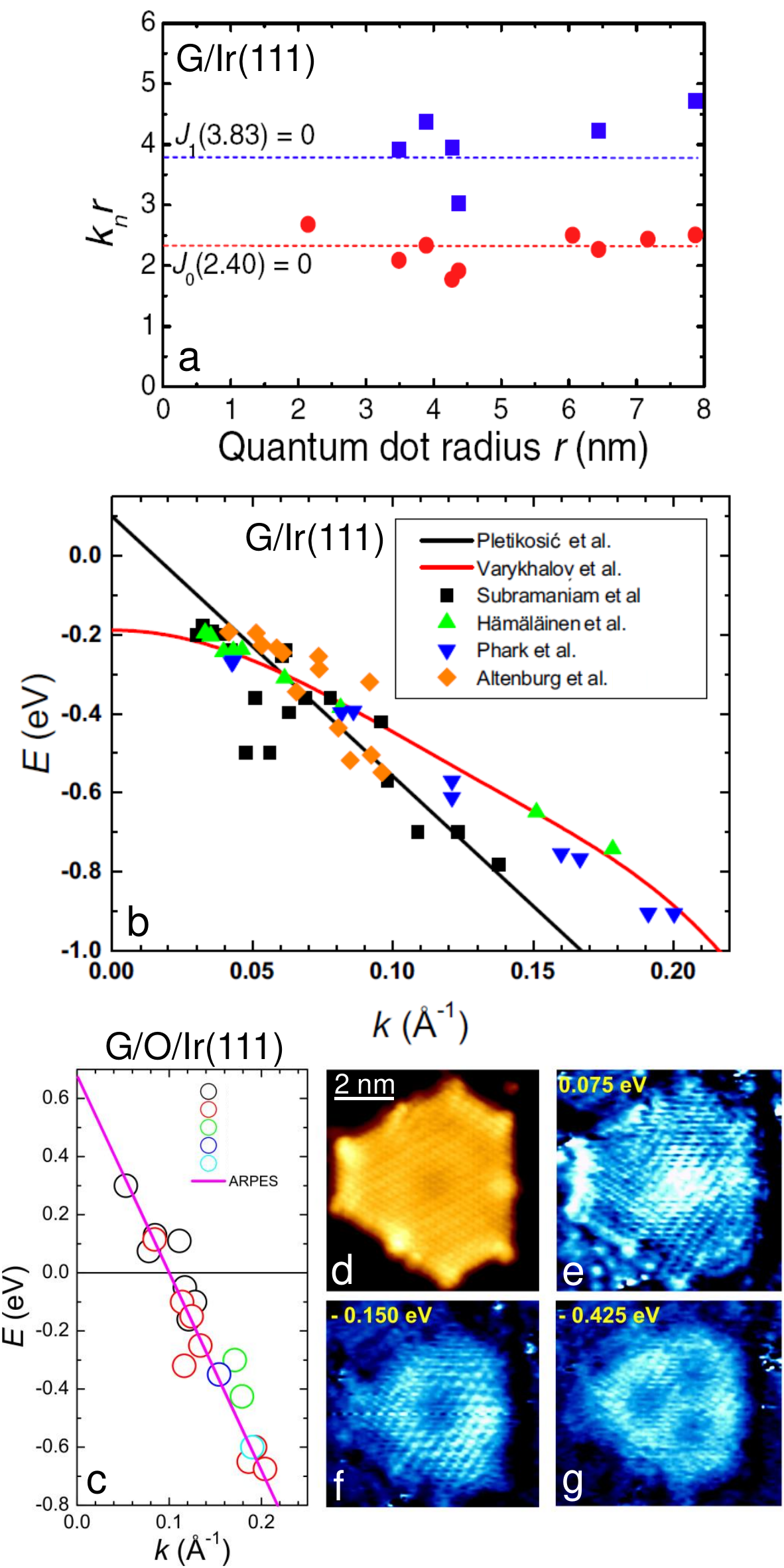}
\caption{(a) Analysis of the QD peak energies in $dI/dV$ curves using a circular model geometry for the graphene QD with radius $r$ determined from the measured QD area $A$ by $r=\sqrt{A}/\pi$; such a model exhibits confined states at $J_n(k_n\cdot r)=0$ with $J_n$ being the n$^{\rm th}$ Bessel function and the graphene dispersion $E_n = \hbar v_{\rm D} k_n$ assuming $v_{\rm D}=1.0\cdot 10^6$ m/s; the agreement between the experimental symbols and the dashed line confirms the Dirac velocity of the confined states to be $v_{\rm D}=1.0\cdot 10^6$ m/s \cite{Subramaniam}; (b) dispersion of confined states in graphene islands on Ir(111): Subramaniam {\it et al.} \cite{Subramaniam} and Altenburg {\it et al.} \cite{Altenburg} deduced $E(k)$ from the method described in (a) except that the measured peak energies $E$ from $dI/dV$ curves and the $k_n$ deduced from $k_n\cdot r = x_{0n}$ with measured QD radius $r=\sqrt{A}/\pi$ and known values of the zeroes of the Bessel functions $x_{0n}$ are used;  H\"am\"al\"ainen {\it et al.} \cite{Hamaleinen} compared the $dI/dV$ images with numerically calculated confined wave functions using a dispersion of Klein-Gordon type, QD state energies are taken to be the $dI/dV$ images with best agreement in spatial distribution; the symbols are then deduced as above using $r=\sqrt{A}/\pi$ with the published QD area $A$; Phark {\it et al.} \cite{Park} used the angularly averaged Fourier transformation of the real space data which exhibits a peak in the intensity vs. $|k_{||}|$ representation taken as the $k$ value; the energy $E$ corresponds to the voltage $V$ where the $dI/dV$ image is recorded according to $E=e\cdot V$ \cite{Jolie}; the lines are deduced from the ARPES data displayed in Fig. \ref{Fig1}(e) (black line, graphene Dirac cone) and Fig. \ref{Fig4}(f) (red line, Ir(111) surface state S$_0$ with graphene on top); (c) same as (b) using the method of Subramaniam {\it et al.} after intercalating 720 L of O$_2$ at $T \simeq 450$ K (symbols) in comparison with dispersion from graphene Dirac cone according to Fig. \ref{Fig1}(e) \cite{Jolie}; (d) STM image of graphene QD on Ir(111) after O$_2$ intercalation; (e)$-$(g) $dI/dV$ images of the graphene QD shown in (d) recorded at the energies $E=e\cdot V$ as marked. \cite{Jolie}
((a) courtesy of F. Libisch, TU Vienna, (b)-(g) courtesy of C. Busse, University of Cologne) 
}
\label{Fig6}
\end{figure*}
The contributions from $S_0$ and from the graphene Dirac cone to the LDOS of the graphene QDs depends on the details of the microtip. Firstly, we show an example where the LDOS is dominated by the graphene states as for the images displayed in Fig. \ref{Fig3}(f)-(h). Figure \ref{Fig6}(a)  shows the first two peak energies deduced from $dI/dV$ curves of the smaller QDs up to an aerial size $A$, which corresponds to an average radius of $r=\sqrt{A}/\pi$. Assuming a circular quantum dot, the analytic solution for confined states is given by the zeroes of the $n^{\rm th}$ Bessel function $J_n$ ($n=0,1,2,...$) according to $J_n(k_n\cdot r)=0$. The wave vector $k_n$ can be taken from the dispersion, which for graphene would read $k_n=E_n/(\hbar v_{\rm D})$ with $E_n$ being the energy of the $n^{\rm th}$ peak in $dI/dV$ curves. Plotting this $k_n$ multiplied by $r$ for different QDs shows good agreement with the zeroes of the Bessel functions $x_{0n}$ ($x_{00}=2.40$, $x_{01}=3.83$), if $v_D=1.0\cdot 10^6$ m/s. This reveals that the particular tips used for these QDs preferentially probe the graphene states, at least, as it concerns the peak structures in $dI/dV$ curves.\\
\begin{figure*}[htb]%
\begin{center}
\includegraphics*[width=0.75\textwidth]{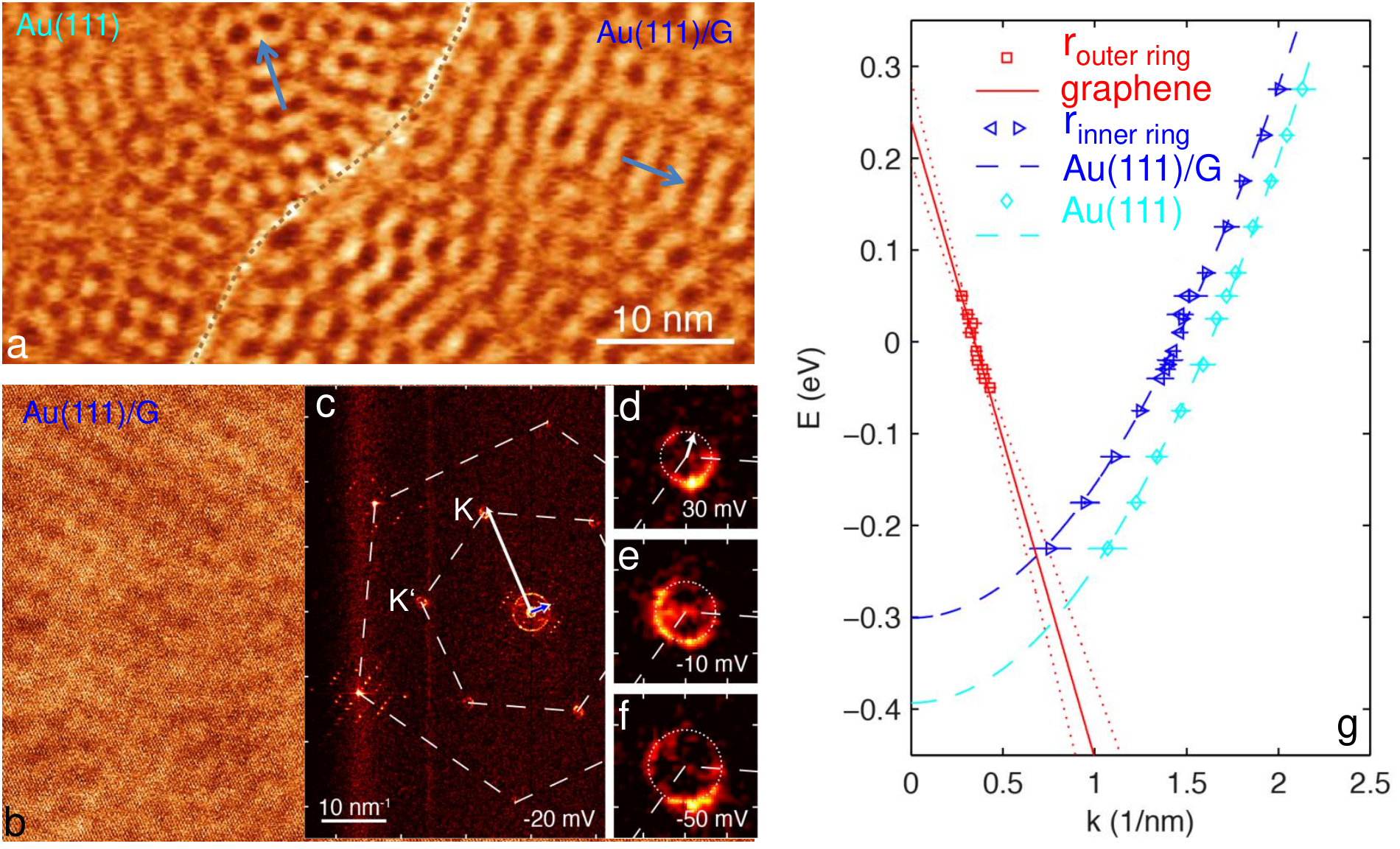}
\end{center}
\caption{(a) $dI/dV$ image of an interface between Au(111) and graphene on Au(111) (G/Au(111)) as marked with the interface highlighted by a dashed line,  $V=-75$ mV, $I=1$ nA, $V_{\rm mod}\simeq 3$ mV; arrows mark directions of wave guiding visible as a one-dimensional orientation of a standing wave \cite{Fujita,Libisch}; (b) $dI/dV$ image on the graphene on Au(111) with atomic resolution, $V=-20$ mV, $I=1$ nA, $V_{\rm mod}\simeq 3$ mV; (c) Fourier transformation (FT) of the $dI/dV$ image in (b) with dashed hexagons marking the Brillouin zone of graphene (smaller hexagon) and the larger reciprocal unit cell connecting the spots which originate from the graphene atoms; white arrow is a vector from $\Gamma$ to a point of the Dirac circle around K, blue arrow marks a vector from $\Gamma$ to a state belonging to the Au(111) surface state; (d)$-$(f) zoom into the area around K at different voltages as marked; (g) ring diameter $k$ deduced from the FT of $dI/dV$ images at different $V=E/e$; red squares: ring around K; dark blue triangles: ring around  $\Gamma$; cyan diamonds: ring around $\Gamma$ from areas not covered by graphene; lines are linear (red) and parabolic (blue,cyan) fits of the equally colored symbols with red dotted lines marking the possible slopes of the red fit line within error bars. \cite{Leicht} (courtesy of M. Fonin, Konstanz University)
}
\label{Fig7}
\end{figure*}
Figure \ref{Fig6}(b) displays the same data as an $E(k)$ dispersion (black squares) using the measured peak energies $E_n$ and $k_n=x_{0n}/r$. The data points fit reasonably with the black line of the graphene Dirac cone as measured by ARPES on Ir(111) \cite{Petikosic,Subramaniam}. However, using the same method except that $r$ is determined from the radius of an inner circle which completely fits into the island, Altenburg {\it et al.} find that their peaks on average fit better to the Ir(111) surface state S$_0$ (see diamonds in Fig. \ref{Fig6}(b)), albeit some of the states could also correspond to the graphene Dirac cone dispersion \cite{Altenburg}. They also provided DFT data showing that an s-type tip being 0.48 ${\rm \AA}$ above the graphene is more susceptible to S$_0$ states than to states from the Dirac cone. In line, they used an Au tip known to be dominated by s-states at $E_{\rm F}$, while Subramaniam {\it et al.} used a W tip, typically showing stronger contributions from d-states. Other groups used different methods to determine $E(k)$ for the graphene QDs on Ir(111). While Phark {\it et al.} (triangles pointing down in Fig. \ref{Fig6}(b)) used the Fourier transformation of the real space $dI/dV$ images resulting in an intensity distribution $I(\underline{k}_{||})$, which displayed six dominant peaks. Angularly  averaging $I(\underline{k}_{||})$ results in $I(|\underline{k}_{||}|)$ and leads to a single peak taken as the $k$ value belonging to the energy $E$ corresponding to the voltage $V=E/e$ of the $dI/dV$ image. The data fit obviously better to the S$_0$ state, albeit not perfectly. Finally, H\"am\"al\"ainen {\it et al.} used a comparison of apparent lateral shapes in calculation and experiment. They visually compared confined squared wave functions in $dI/dV$ images and confined squared wave functions from numerical calculations using the Klein-Gordon equation. The Dirac velocity $v_{\rm D}$ and, thus, the wave vector $k$ at given $E$ is adapted in the calculation until it fits to the experimentally found energy, where the calculated state resembles the experimental one most strongly (upwards triangles in Fig. \ref{Fig6}(b)). This comparison again results in a better fit of $E(k)$ to S$_0$ than to the Dirac cone of graphene (Note that the claim in the original publication was different \cite{Hamaleinen}).\\
It somehow appears that, firstly, the shapes of the confined wave functions (used in \cite{Park} and \cite{Hamaleinen}) are more strongly influenced by S$_0$ than the peaks in $dI/dV$ curves (used in \cite{Subramaniam} and \cite{Altenburg}). This points to a longer lifetime of the Dirac cone states than of the Ir(111) S$_0$ state, which makes the peaks belonging to the Dirac cone sharper than the peaks belonging to S$_0$. This suspicion appears to be in line with the different energetic widths of S$_0$ and the Dirac cone found in the ARPES data of Fig. \ref{Fig1}(e) and Fig. \ref{Fig4}(f). However, secondly, the relative importance of the two contributions can depend on details of the tip which are not under control in the STM measurements.\\

An easy way to get rid of the S$_0$ contribution is the intercalation of oxygen (750 L O$_2$ at 450 K) between the graphene QD and the Ir(111) \cite{Jolie}. This, firstly, removes S$_0$  as evidenced by ARPES \cite{Larciprete} and, secondly, increases the graphene-Ir(111) distance. Indeed, Fig. \ref{Fig6}(c) shows that the resulting $E(k)$ dispersion deduced from the peak energies and the island sizes, as described for Subramaniam {\it et al.} above, fits rather nicely to the graphene Dirac cone, which is additionally p-doped by the oxygen as also deduced by ARPES \cite{Jolie}. The corresponding confined wave functions, shown in Fig. \ref{Fig6}(e)-(g), exhibit the same regular spatial appearance as discussed in Fig. \ref{Fig3} indicating that the soft confinement driven by the edge-Ir(111) interaction is still present even after oxygen intercalation.\\

While this line of probing of graphene QD states has not been pursued further so far, an interesting way to distinguish the contributions from Dirac cone states
and interfering surface states has been published \cite{Leicht}. In order to decouple the graphene QDs from the Ir(111), the authors intercalated 5-10 nm of Au(111) between the QDs and Ir(111) by Au evaporation at
300 K. Indeed, the graphene islands on top of Au(111) could now easily be moved by the forces of the tip of the STM indicating that strong chemical bonds as between Ir(111) and the graphene edges are not present on Au(111). Nevertheless, the dominating Au(111) surface state, being located around $\Gamma$ and intersecting $E_{\rm F}$ \cite{Keven,LaShell}, still interferes with the Dirac cone states. Figure \ref{Fig7}(a) shows a $dI/dV$ image with an interface (dashed line) between uncovered Au(111) areas (left) and  Au(111) areas covered with graphene (right). Both areas show very similar standing wave patterns. Interestingly, the wave guiding along the herringbone reconstruction (arrows) \cite{Fujita}, recently explained as a diffraction effect \cite{Libisch}, is visible, too, on both sides of the interface. Thus, obviously the Au(111) surface state, known to have a parabolic dispersion as $E(k)=\hbar^2 k^2/(2m^\star)+E_{0}$ with an origin at $E_0-E_{\rm F}=-480$ meV and an effective mass of $m^\star=0.25\cdot m_{\rm e}$ ($m_{\rm e}$: electron mass) \cite{Keven,LaShell}, is visible by tunneling on the graphene revealing that it is penetrating into the graphene like the S$_0$ state of Ir(111).\\
However, by recording $dI/dV$ images with atomic resolution as shown in Fig. \ref{Fig7}(b), it gets clear that the graphene Dirac cone contributes to the observed standing wave patterns as well. The Fourier transformation (FT) of the real-space $dI/dV$ image (Fig. \ref{Fig7}(c)) shows a central circle compatible with the Au(111) surface state and, in addition, circles which correspond to intervalley scattering between K and K', i.e. between the Dirac cones of graphene. These circles show an energy dependence of its diameter as expected for the graphene Dirac cone (Fig. \ref{Fig7}(d)-(f)), i.e.,    while the inner circle belonging to the Au(111) surface state gets larger with increasing energy, the outer circle gets smaller. Plotting the diameter of the circles as a function of energy (Fig. \ref{Fig7}(g)) allows to deduce the dispersion of the contributing bands. The authors find a parabolic dispersion for the inner circle
with an effective mass fitted to be $m^\star=0.26\pm 0.02\cdot m_{\rm e}$ as found identically on the uncovered Au(111) areas. This identifies this inner ring as being caused by the Au(111) surface state penetrating into the graphene. The fitted band onsets, $E_0-E_{\rm F}=-390$ meV on clean Au(111) and $E_0-E_{\rm F}=-300$ meV on graphene covered Au(111), are higher than on the Au(111) surface of a single crystal, which is probably due to confinement effects of the thin Au(111) layer on the Ir(111) and charge transfer from the graphene, respectively. The diameter of the outer rings reveals a linear dispersion $E=\hbar v_{\rm D} k +E_{\rm D}$ with fit parameters $v_D=(1.1\pm 0.2)\cdot10^6$ m/s and $E_{\rm D}=0.24\pm 0.05$ eV identifying it as belonging to the graphene Dirac cone. Thus, in principle, the intensity ratio between the inner and the outer rings provides access to the relative contribution of Dirac cone states and surface states of the support to the $dI/dV$ images of the graphene QDs. It also allows to discriminate their spatial shapes by an adequately filtered back transformation of the FT data, an interesting experiment still to be done.\\

\subsection{Edge states}
\begin{figure*}[htb]%
\begin{center}
\includegraphics*[width=1.0\textwidth]{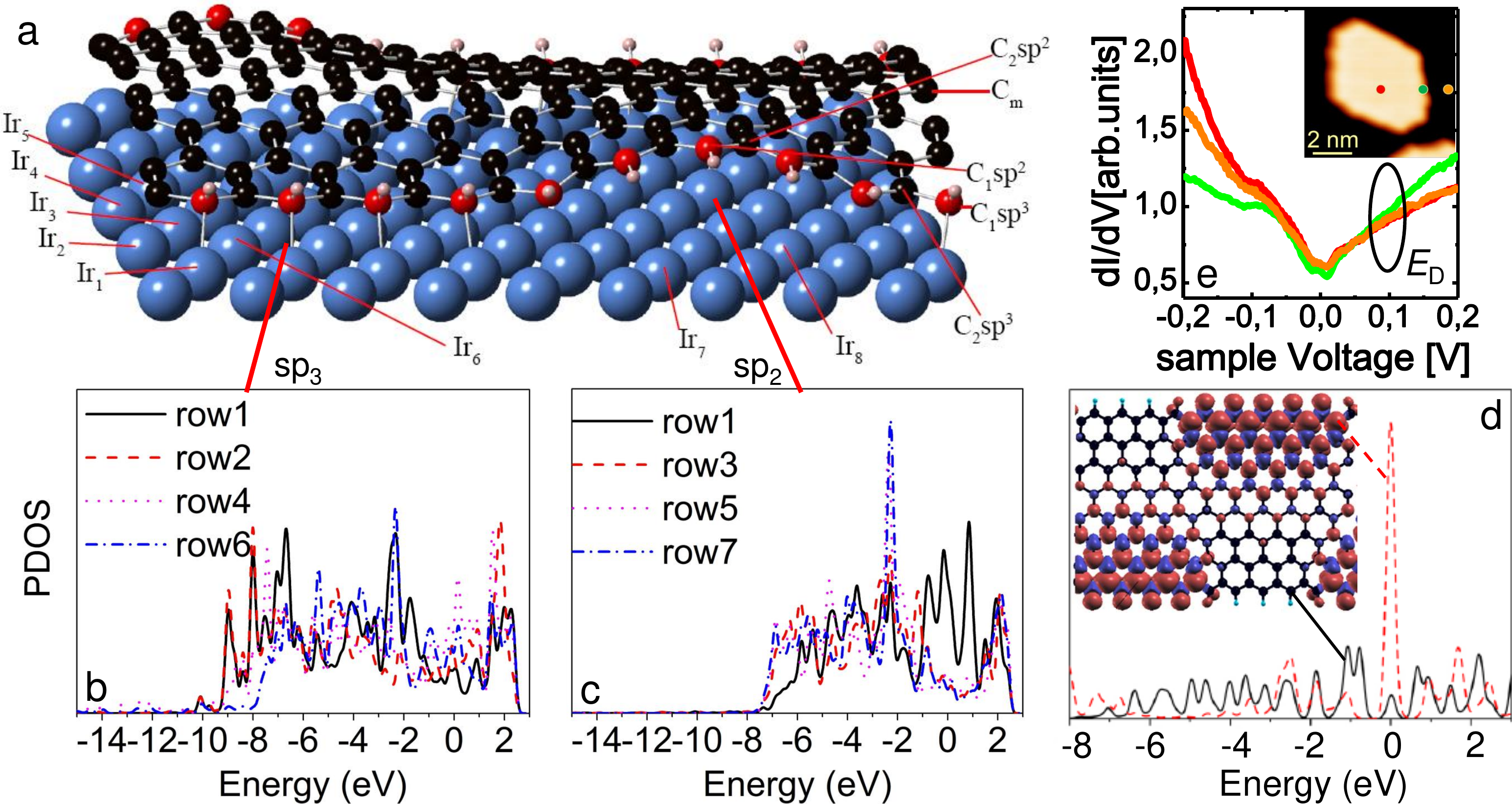}
\end{center}
\caption{(a) Atomic arrangement of a graphene nanoribbon on the Ir(111) substrate (blue balls: Ir atoms, black balls: C atoms of graphene bulk, red balls: graphene edge atoms, pink balls: H atoms); the arrangement is deduced from the relaxation within a DFT calculation; for the sake of clarity, only the top Ir layer is shown, while four layers of Ir are used in the calculations; different atoms are numbered and the bonding configuration (sp$^2$ or sp$^3$) of several C atoms at the edge is marked; (b)-(c) projected density of states (PDOS) of particular atomic spheres; (b) atoms behind the sp$^3$ bond edge atom as indicated by the red line; rows are numbered away from the edge; (c) same as (b) for the atoms behind the sp$^2$-type edge atom marked by a red line, too; (d) PDOS of C edge atoms (as marked by lines) of a freely suspended graphene nanoribbon as sketched in the inset with partial twofold H-termination (colored areas) and partial one-fold H-termination (cyan H atoms at black C atoms); the red and blue colors in the inset mark the oppositely oriented spin polarizations of the atoms with the size of the colored ball marking the strength of spin polarization; (e) $dI/dV$ curves recorded at the positions marked in the STM images of a graphene QD in the inset, $I_{\rm stab}=0.3$ nA, $V_{\rm stab} = 0.3$ V, $V_{\rm mod} = 5$ mV; the area around $E_{\rm D}$, where an edge state would be expected, is marked. \cite{Li2}
((a)$-$(d) courtesy of R. Mazzarello, RWTH Aachen University)
}
\label{Fig8}
\end{figure*}
Coming back to the graphene QDs on Ir(111), we discuss its edge properties. As shown in Fig. \ref{Fig1}(b), the edge is completely zigzag type. This might be regarded as favorable considering that the armchair edge is much more stable for freely suspended graphene \cite{Wassmann,Kotakoski} and that graphene zigzag edges are predicted to exhibit an edge state becoming ferromagnetic if $E_{\rm D}\simeq E_{\rm F}$ \cite{Yamashiro,Pisani}. However, this edge state is suppressed on Ir(111) as will be explained in the following.\\ 
DFT calculations of a graphene ribbon on top of Ir(111) show that the strong interaction between the Ir d$_{z^2}$-orbitals and the p$_z$ orbitals of graphene lead to a preferential on-top position of the C-atoms at the edge with respect to Ir atoms (see relaxed geometry in Fig. \ref{Fig8}(a), left area). The bond induced downwards bending of the C atoms by about 1.6 ${\rm \AA}$ leads to a sp$^3$ like bonding configuration of the C atom with two bonds to neighboring C's and one bond to the neighboring Ir d$_{z^2}$ orbital. The remaining sp$^3$ bond is most likely saturated by hydrogen which originates from the preparation via C$_2$H$_4$. The latter has been verified by comparison between measured STM images at different bias and simulated STM images based on DFT \cite{Li2}. Moreover, the single-H terminated C-atoms at the edge are energetically favorable with respect to double-hydrogenated C-atoms and unsaturated C atoms on top of Ir(111) \cite{Li2}. Importantly, the preferential on-top position of the C-atoms causes strain within the graphene edge region due to an effective lattice mismatch between graphene and Ir(111) of about 10 \%. This, in turn, causes strain relaxation areas, where the graphene edge is more detached from the Ir(111) being close to its natural distance in the interior of extended graphene of $3.4\pm 0.2$ ${\rm \AA}$ \cite{Busse} (see right area of Fig. \ref{Fig8}(a)). These more detached areas mimic a sp$^2$ bonding configuration with two bonds connected to neighboring C atoms and the remaining one being saturated by H, while the edge p$_z$ orbital remains unsaturated.\\
Surprisingly, the DFT calculations did not find indications of the predicted edge state being a hallmark of the zigzag edge. This is shown by the projected density of states (PDOS) of different atoms around the edge (Fig. \ref{Fig8}(b) and (c)), which do not show very pronounced features in the area around $E_{\rm F}$, i.e. at $E=0$ eV. While the C atoms in the sp$^3$ areas show a suppressed PDOS with respect to the surroundings, the atoms from the sp$^2$ area show a triple peak at the very last atom, which is, however, not larger than other peaks of the PDOS. For comparison, the red curve in Fig. \ref{Fig8}(d) shows a typical edge state which is a factor of $5-10$ higher than all surrounding peaks. In line, the $dI/dV$ curves recorded at the edge of the graphene QDs on Ir(111) do not show any spectral feature around $E_{\rm D}$ as marked in Fig. \ref{Fig8}(e). This result is found for several tens of graphene QDs using  a similar number of different microtips.\\
The absence of the edge state is surprising, since it is generally believed that the edge is a relatively robust feature of graphene zigzag edges. It originates from the fact that the introduction of radical $\pi$-bonds on every third edge atom allows the formation of Clar's sextets (six double bonds in one benzoid ring) through the whole interior of the graphene bulk, such that substantial kinetic energy by the corresponding delocalization of electrons within the p$_z$ orbitals of the ring can be gained \cite{Wassmann}. The radicals at the edge then form the flat band at $E_{\rm D}$ being prone to a ferromagnetic instability, if $E_{\rm D}=E_{\rm F}$ \cite{Yazyev}. Thus, at first hand, the soft confinement found in the previous paragraph for graphene QDs on Ir(111) would just move the necessity of these radicals to the interior of the graphene and, thus, would lead to an edge state away from the physical edge of the graphene QD. Such an edge state, however, is also not observed neither in DFT nor in STS.\\
There are two possibilities to explain the disappearance of the edge state. Firstly, the bonding along the edge is not translationally invariant on the atomic scale, but resembles a zz(2222222111) configuration in the nomenclature by Wassmann et al. \cite{Wassmann} where zz symbolizes the zigzag edge and the numbers describe the number of saturated bonds in the periodically repeated sequence of ten edge atoms. Our DFT calculations of a freely suspended graphene ribbon with such an edge termination, represented by H atoms, reveal that this termination, indeed, partly destroys the edge state, i.e. the radicals are only required in the area of the double bonds in order to allow the formation of Clar's sextets \cite{Li2}. This can be rationalized by the fact that a zigzag edge length of three unit cells can host Clar's sextets in the whole interior of the graphene \cite{Wassmann}. However, for an extended edge, this requires that the surrounding edge areas host ionized p$_z$ bonds to adapt
Clar's sextets in the rest of the graphene bulk. Secondly, the presence of the substrate can prefer certain locations of double bonds by the interaction with the d$_{z^2}$ orbital of the Ir(111), such that the gain in kinetic energy for Clar's sextets is overcompensated by the exchange energy between Ir d$_{z^2}$ orbitals and the C p$_z$ orbitals. Such an effect should be directly visible in STM images which would show these preferred bonds as occupied states close to $E_{\rm F}$, while the unfavorable bonds are visible as unoccupied states \cite{Wassmann}.

\begin{figure}[htb]%
\includegraphics*[width=0.5\textwidth]{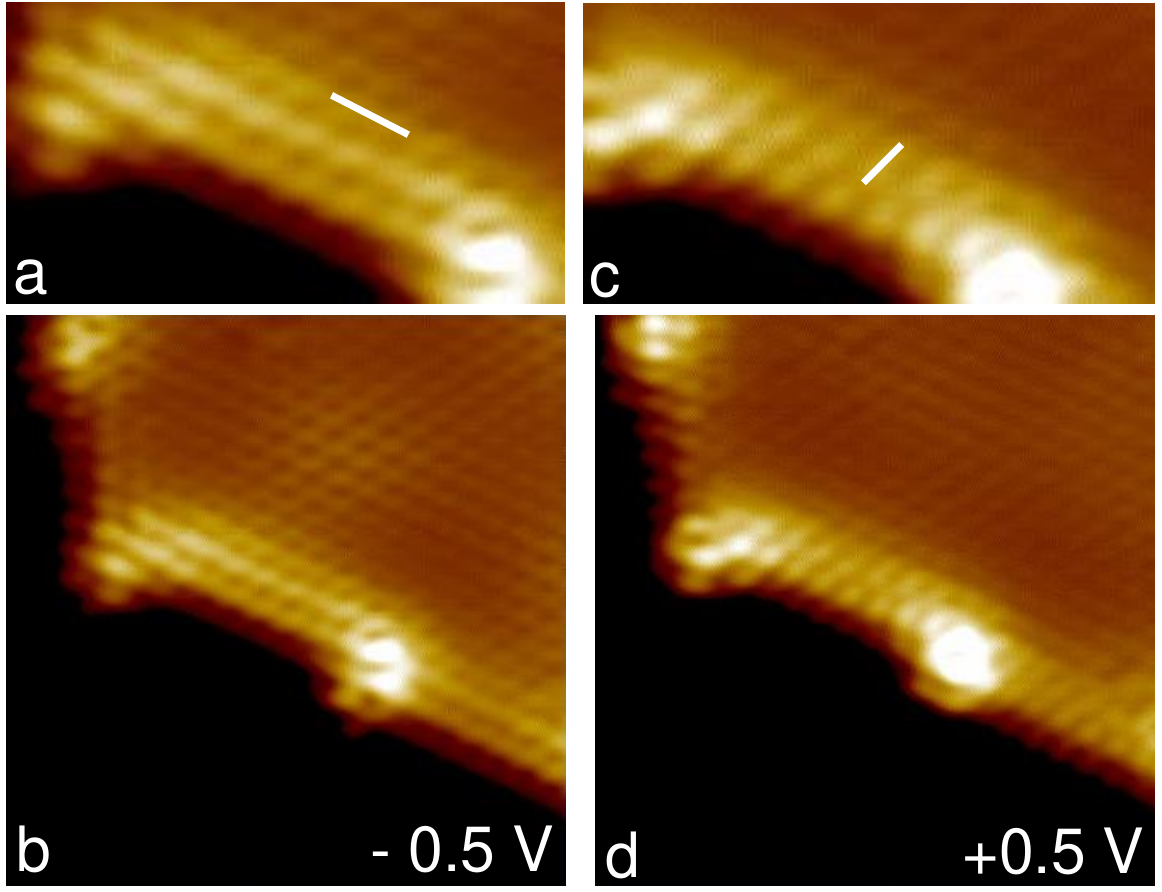}
\caption{(a)$-$(d) STM-image of edge of graphene QD on Ir(111) with atomic resolution; (a), (b) $V=-0.5$ V, $I=5$ nA; (c), (d) $V=-0.5$ V, $I=5$ nA; white lines mark preferential bond orientations within the edge regions of the images.
}
\label{Fig9}
\end{figure}
Indeed, differences between brightly appearing bonds in occupied and empty state STM images (negative and positive bias, respectively) are visible in Fig. \ref{Fig9}. 
While the bonds perpendicular to the zigzag edge are preferentially empty, other bonds more parallel to the zigzag edge appear to be occupied. 
Thus, we conclude that the lattice mismatch between Ir(111) and graphene, leading to the zz(2222222111) configuration and to the inequivalence between C-C bonds on the substrate in combination with the d$_{z^2}$ orbital at $E_{\rm F}$, leading to a relatively strong interaction with the graphene p$_z$ orbital, suppresses the edge state on Ir(111) \cite{Li2}. This is favorable for graphene QDs to be used as spin qubits \cite{Trauzettel}, since the edge magnetism could interact with the qubit spin, thus, offering a prominent channel for decoherence. However, it is not straightforward to transfer this concept of edge state suppression to graphene QDs on insulating substrates.\\
\begin{table}[hbt]
  \caption{DFT results regarding the presence of magnetic edge states at graphene zigzag edges on different metallic substrates with different terminations \cite{Li3}.}
  \begin{tabular}{c c c c}
   \hline
    substrate/ & edge state & magnetic & reason \\
    termination & & &\\
    \hline
    Ir(111)/H  & No  & - & hybrid with Ir 5d$_{z^2}$  \\
    Ir(111)/bare  & No  & - & hybrid with Ir 5d$_{z^2}$\\
    Au(111)/H  & Yes  & Yes & -  \\
    Au(111)/bare  & Yes  & No & charge transfer to Au\\
    Ag(111)/H  & Yes  & No & charge transfer to Ag  \\
    Ag(111)/bare  & Yes  & No & charge transfer to Ag\\
    Cu(111)/H  & Yes  & No & charge transfer to Cu \\
   Cu(111)/bare  & Yes  & No & charge transfer to Cu\\
    \hline
  \end{tabular}
  \label{tab1}
\end{table}

In order to demonstrate that the suppression of the edge state is rather the exception than the rule, we performed calculations of graphene nanoribbons on different close-packed metallic surfaces such as Ag(111), Cu(111) and Au(111), either with single H-termination of the edge or without termination \cite{Li3}. The results in combination with the ones on Ir(111) are shown in table \ref{tab1}. All zigzag edges except the ones on Ir(111) exhibit an edge state, most likely, since none of these surfaces has a d$_{z^2}$ surface state close to $E_{\rm D}$. Thus, the presence of this particular surface state pointing directly to the graphene p$_z$ orbital and being partly occupied, since at $E_{\rm F}$, is decisive for the suppression of the edge state as analyzed above.\\
Albeit most of these substrates leave the edge state intact, the charge transfer from the edge states to the substrate mostly results in an empty edge state, which is not prone to a ferromagnetic instability, as consistently found in the DFT calculations \cite{Li3}. Only a Au(111) substrate, which exhibits the weakest interaction and only after an additional H-termination of the graphene edges, which further reduces the interaction with the Au(111) d$_{z^2}$ surface states being located 2 eV below $E_{\rm F}$, leaves the charge transfer small enough, such that a ferromagnetic instability occurs according to DFT \cite{Li2}. In line, the published indirect evidence for a ferromagnetic edge state \cite{Tao} was observed on Au(111), however, without a controlled edge termination. Subsequent hydrogenation of the corresponding graphene nanoribbon edges has also been achieved \cite{Zhang2}, however, so far, without an investigation of the magnetic properties of the edge.\\
\begin{figure}[tbh]%
\includegraphics*[width=0.5\textwidth]{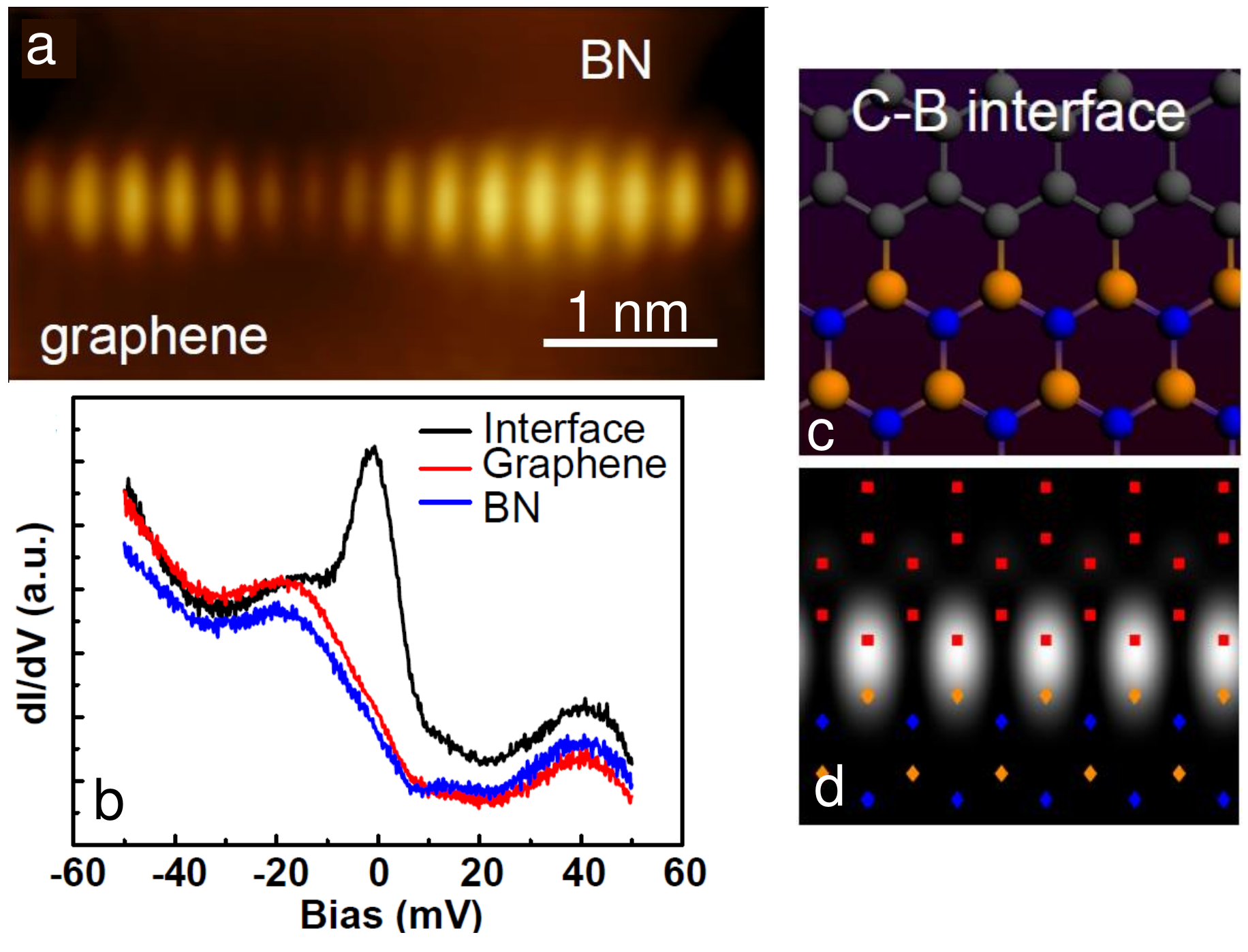}
\caption{(a) STM-image of the lateral zigzag interface between graphene and BN as marked on Au(111)/Ir(111) showing the edge state as a bright area, $V=-1$ V, $I=3$ nA; (b) $dI/dV$ curves recorded on graphene, BN, and the lateral interface as marked;  (c) stick-and-ball model of a freely suspended lateral graphene-BN interface with C-atoms (grey), B-atoms (orange), and N-atoms (blue); (d) DFT based simulation of the spatial distribution of the tunneling current of the assembly of (c) at $V=-0.2$ V with positions of C-atoms (red), B-atoms (orange), and N-atoms (blue) overlaid. \cite{Drost} (courtesy of P. Liljeroth, Aalto University)
}
\label{Fig10}
\end{figure}
Also in line, the edge state of the graphene QDs on Ir(111) can be recovered by intercalating Au(111). Therefore, Drost {\it et al.} \cite{Drost}, firstly, terminated the graphene zigzag edges by BN, which can be grown on Ir(111) using borazine ((HBNH)$_3$) \cite{Preobrajenski} and which seamlessly contacts the previously prepared graphene QD edges along the zigzag directions. Secondly, they intercalated Au such that the interaction with the substrate is additionally reduced by the removal of the d$_{z^2}$ surface state of Ir(111). Figure \ref{Fig10}(a) shows a STM image of the lateral  interface between graphene and BN on the Au(111) showing a state at the interface which slightly varies in perpendicular size probably due to a remaining influence of the substrate. The corresponding state could be identified by $dI/dV$ curves (FIG. \ref{Fig10}(b)). It is found to be within $\pm 5$ meV around $E_{\rm F}$ in accordance with DFT calculations \cite{Drost}. It is not clear so far, if the state is ferromagnetic, which can be probed, e.g., by spin-polarized STM \cite{Bode}. However, the lateral distribution of the state can be nicely reproduced by DFT ignoring the substrate as shown in Fig. \ref{Fig10}(c) and (d), if the energetically preferential B-termination of the BN-graphene  zigzag interface is chosen \cite{Drost}.\\
Thus, an edge state can indeed be recovered from the favorable production of zigzag edges on Ir(111) being located close to $E_{\rm F}$. It might exhibit the long-sought edge magnetism directly. This would be not favorable for the spin qubit applications, but other spintronic applications as, e.g., spin filters relying on this edge state have been proposed \cite{Son}.

\section{Anodically oxidized quantum dots}
\begin{figure*}[htb]%
\begin{center}
\includegraphics*[width=1.0\textwidth]{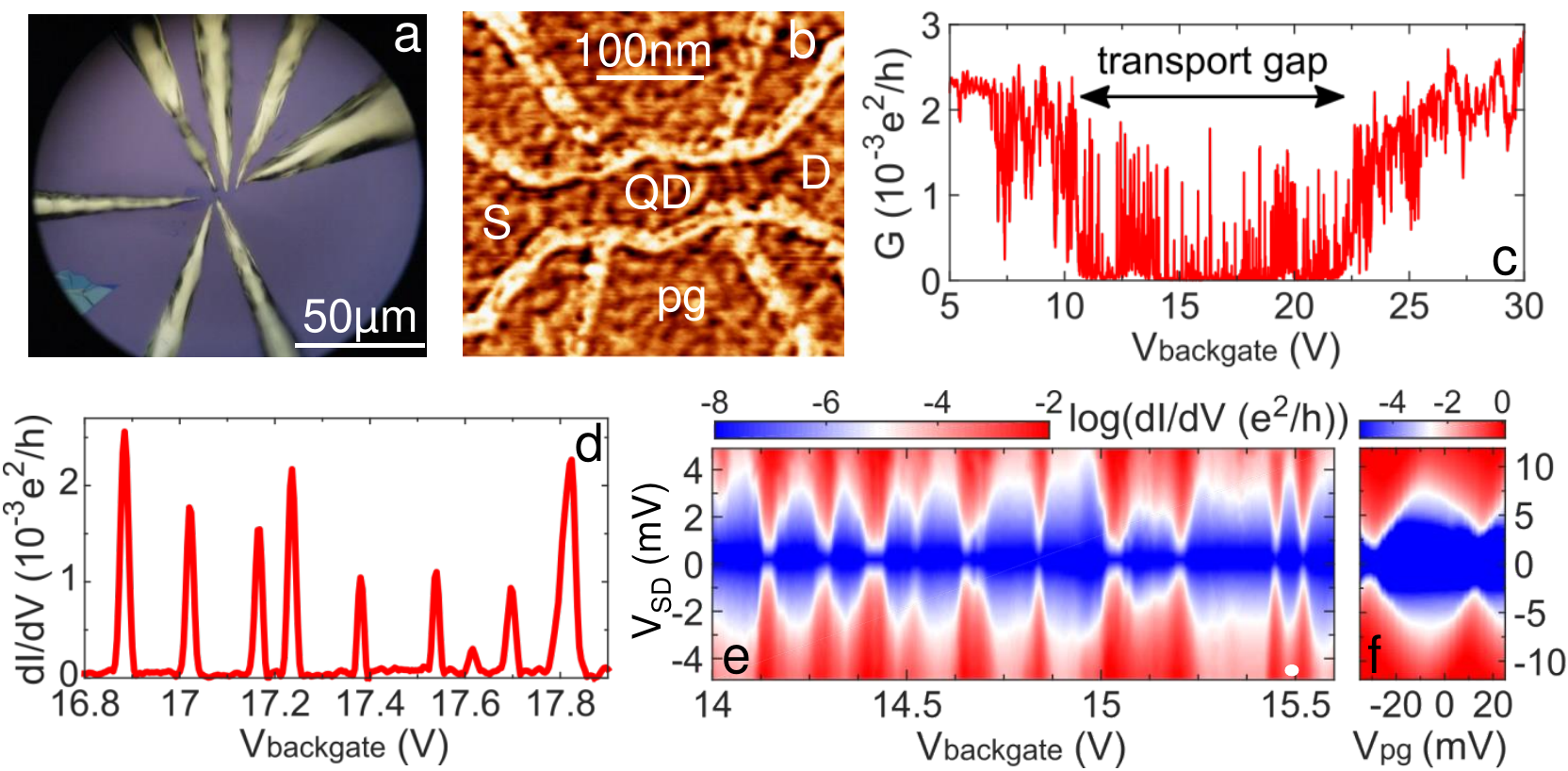}
\end{center}
\caption{(a) Optical micrograph of a graphene flake on Si/SiO$_2$ contacted by seven In lines microsoldered onto the graphene flake \cite{Geringer2}; (b) AFM tapping-mode image (phase image) of a graphene QD (QD) on Si/SiO$_2$ with source (S) and drain (D) contacts as well as a number of plunger gates; the plunger gate used in (f) is marked as pg; the oxidized lines (bright lines) are produced by LAO in AFM contact mode at a controlled humidity of 60 \%, $V_{\rm tip}=-8.5$ V, $v_{\rm tip}=200$ nm/sec; (c) conductance $G$ as a function of $V_{\rm backgate}$ at $V_{\rm SD}=1.5$ mV and $V_{\rm pg}=-20$ mV; all other contacts are grounded, $T=0.3$ K; transport gap where $G$ partly drops to zero is marked; (d) $dI_{\rm SD}/dV_{\rm SD}$ as a function of $V_{\rm backgate}$ within the transport gap region; an oscillating $V_{\rm SD}$ with amplitude $V_{\rm SD_{max}}=0.1$ mV is used for lock-in detection; all other contacts are grounded, $T=1.2$ K; (e) logarithmic representation of $dI_{\rm SD}/dV_{\rm SD}$ as a function of $V_{\rm SD}$ and $V_{\rm backgate}$ with all other contacts grounded; an oscillating voltage with $V_{\rm SD_{max}}=0.1$ mV is added to $V_{\rm SD}$ for lock-in detection, $T=1.3$ K; the white dot marks a well-developed Coulomb diamond discussed in the text; (f) logarithmic representation of $dI_{\rm SD}/dV_{\rm SD}$ as a function of $V_{\rm SD}$ and $V_{\rm pg}$ with $V_{\rm backgate}=14.8$ V and all other contacts grounded; an oscillating voltage with $V_{\rm SD_{max}}=0.03$ mV is added to $V_{\rm SD}$ for lock-in detection, $T=0.3$ K.
}
\label{Fig11}
\end{figure*}
The described STM experiments on graphene QDs deposited on metals provide an unprecedented spatial resolution concerning the electronic properties of the QDs, but 
they are difficult to be directly correlated to transport properties of graphene QDs, which are typically measured on insulating substrates and which are defined in a top-down approach by etching and subsequent contacting of graphene flake structures using e-beam lithography \cite{Guttinger}. It is believed that the difficulties to control the QD transport properties in a similar way as in GaAs \cite{Kouwenhoven} is mostly due to the uncontrolled edge geometry and chemistry \cite{Bischof,Libisch2}. This is distinct to the QDs studied in UHV by STM, where these two parameters are well controlled.\\
In order to locally probe transport QDs, scanning gate microscopy has been used, which, however, suffers from a rather limited spatial resolution of about 100 nm \cite{Schnez,Conolly}.
STS has a much higher spatial resolution, but requires that the QD is on the surface, since the tunneling current decays by an order of magnitude, if the tip-QD distance is increased by merely 1 ${\rm \AA}$ \cite{Wiesendanger}. Moreover, the surface of the QD has to be free of insulating resist, which otherwise would be charged by the electric field of the tip. In turn, this charging of resist can change the local potential below the tip and, thus, the LDOS contributing to the tunneling current \cite{Wildoer}.
Consequently, it is mandatory to structure and contact the graphene nanostructures without resist.\\
For contacting, one can either use shadow mask evaporation \cite{Staley} or contacting by microsoldering \cite{Girit}. We used the second method \cite{Geringer2}, which basically melts a fine wire of In onto the graphene sample held at about 155$^o$ C. The In wire is pulled out of an In droplet by a typical STM tip made out of W. The positioning of this In wire onto the sample is afterwards done with micromanipulators under an optical microscope. This leads to a multitude of In contacts placed with a precision of 1-10 ${\rm \mu}$m as shown in Fig. \ref{Fig11}(a). In order to structure the graphene into a QD, we used local anodic oxidation (LAO) by the tip of an AFM \cite{Campbell}. This technique has successfully been used previously to produce nanostructures in GaAs \cite{Held}. It also works on graphene \cite{Weng}, where, e.g., quantum dots \cite{Neubeck} or constrictions \cite{Masubushi} made by LAO have been probed by transport spectroscopy.\\
The basic technique is to move the cantilever of an AFM across the surface with a velocity $v_{\rm tip} \simeq 100$ nm/s while applying a negative tip bias $V_{\rm tip}$. A controlled humidity leads to an adhesive droplet of water between tip and sample, where the electric field between tip and sample dissociates the H$_2$O and guides the oxygen radical to the sample surface, such that an oxidation of the graphene can take place.
The C-oxidation on graphene can be either complete leading to cuts in the graphene \cite{Weng} or incomplete, most likely leading to partially oxidized graphene nanoislands, which provide insulating layers, too \cite{Byun,Masubushi2}. These partially oxidized areas appear elevated with respect to the intact graphene in tapping mode AFM images. They are electrically insulating up to voltages of about 2 V \cite{Masubushi2}. Typically, cuts are surrounded by incompletely oxidized areas. Reducing the tip voltage, in order to selectively produce the oxidized nanoislands, leads to the laterally smallest insulating barriers.\\
As shown in Fig. \ref{Fig11}(b), we have produced a graphene QD with a diameter of about 60 nm using LAO. This QD contains source (S) and drain (D) contacts and a number of 
plunger gates (pg) in order to tune the electrostatic potential of the QD and its tunneling barriers. The QD is placed on a Si/SiO$_2$ substrate such that the highly doped Si can act as a backgate. We could show that the etching process of graphene by LAO also works on BN supports.\\
Figure \ref{Fig11}(c) shows the measured 2-point conductance $G=I_{\rm SD}/V_{\rm SD}$ ($V_{SD}$: voltage applied between S and D, $I_{\rm SD}$: current measured in the circuit including S, D, and QD) through the QD as a function of the backgate voltage $V_{\rm backgate}$ revealing a $V_{\rm backgate}$ area, where the conductance in between peaks drops towards zero as typical for Coulomb blockade \cite{Kouwenhoven}. Zooming into this $V_{\rm backgate}$ area exhibits that the differential conductance $dI_{\rm SD}/dV_{\rm SD}$ indeed drops significantly between peaks of full width at half maximum (FWHM) of $\Delta V_{\rm backgate}=15$ mV. Plotting the differential conductance as a function of $V_{\rm backgate}$ and $V_{\rm SD}$ reveals the typical Coulomb blockade structures (Fig. \ref{Fig11}(e)) known for QDs \cite{Kouwenhoven,Stampfer,Ponamerenko}. However, the patterns in Fig. \ref{Fig11}(e) appear like overlapped Coulomb structures from QDs connected in series.
Only a few of the diamonds like the one marked by a white dot  have the characteristic shape of an isolated charging event and can be used to determine, e.g., the lever arm of the back-gate voltage to the QD to be $\alpha = 0.067$. Most likely, our structure still suffers from additional confinement areas within the source and drain regions, which produce additional quantum dots in series, as partly observed in etched graphene QDs, too \cite{Stampfer2}. The size of the marked Coulomb diamond fits reasonably to the QD area of Fig. \ref{Fig11}(b). Taking a QD area of $A=60 \times 200$ nm$^2$, a thickness of the SiO$_2$ of $d=300$ nm and a dielectric constant of $\epsilon =4$, we get an estimated back-gate-QD capacitance $C_{\rm bg}$ using the plate capacitor model:
\begin{equation}
C_{\rm bg}=\frac{\epsilon \cdot \epsilon_0 \cdot A}{d} = 1.4 \hspace{2mm} {\rm aF}.
\end{equation}
Dividing this by the lever arm $\alpha$ results in a total capacitance of $C_{\rm tot}=C_{\rm bg}/\alpha=21$ aF such that we get a charging voltage of 
$\Delta U =e/C_{\rm tot}= 7.6$ mV in reasonable correspondence to the 4 mV found for the marked Coulomb diamond in Fig. \ref{Fig11}(e). The remaining discrepancy is probably due to the inadequate plate capacitor model, since the size of the QD is smaller than its distance to the backgate leading, e.g., to fringe field amplification \cite{Ihn}.\\
Fig. \ref{Fig11}(f) shows that the graphene QD can also be tuned by the plunger gate with a much better lever arm close to $\alpha = 0.2$, which is a typical value of $\alpha$ as also found for etched graphene QDs \cite{Stampfer,Ponamerenko, Guttinger,Stampfer2,Volk}. Using the charging voltage of the plunger gate to estimate its capacitance, we get $C_{\rm pg}= e/\Delta U_{\rm pg}\simeq 3.5$ aF ($\Delta U_{\rm pg}\simeq 45$ mV) such that the six gates as well as source and drain can easily account for the lever arm of the back gate.\\
Thus, the anodic oxidation can produce graphene QDs tunable by backgate and sidegates, where the latter exhibit and extremely strong tuning ability. Moreover, the QDs are produced without any resist, such that they can be probed by transport spectroscopy and scanning tunneling spectroscopy at the same time. A scanning tunneling microscope having five transport contacts for the sample, operating in UHV at 0.4 K, and in magnetic fields up to 14 T is available \cite{Bindel}, such that the mentioned experiments to map the quantum dot states responsible for transport are within reach. However, it would be favorable to probe quantum dots with isolated Coulomb diamonds only, using, e.g., a BN support \cite{Engels}.

\section{Conclusion}
We have shown, that scanning tunneling spectroscopy of graphene nano-islands on metallic substrates, in particular on Ir(111) and Au(111), can probe the squared wave functions of graphene QDs, which are, however, intermixed with metallic surface states penetrating into the graphene layer. Effective ways to decouple the surface states are either, physically, by intercalating oxygen between graphene and the metal or, analytically, by performing a Fourier transformation provided that the real space images exhibit atomic resolution, such that the contributions from graphene states located at K and K' and the metallic surface states located at $\Gamma$ get distinct in Fourier space. First investigations of the graphene QD states on Ir(111) showed that the soft confinement induced by the interaction between the graphene p$_z$ orbital and the Ir d$_{z^2}$ orbital at $E_{\rm F}$ leads to rather regular confined states, which are not prone to a detrimental K-K' mixing by the edge of the QD. Moreover, this interaction in combination with the 10\% lattice mismatch between Ir(111) and graphene suppresses the edge state, otherwise, typical for zigzag edges of graphene, such that it can not interfere with the spin properties of the confined state. In contrast, the edge state is predicted to be present, e.g., on Cu(111), Ag(111) and Au(111) supports and it can be recovered by laterally interfacing the zigzag edges on Ir(111) with BN and subsequently intercalating Au. Thus, a tunability of the graphene QDs in a favorable or unfavorable way by the interaction with the substrate  and/or the lateral interfaces has been demonstrated, which might be a helpful guide to tune quantum dot properties also for applications.\\
In order to go beyond these model studies of graphene QDs under well controlled conditions, we finally presented an approach to combine transport studies and scanning tunneling spectroscopy directly. Therefore, we use local anodic oxidation for nanostructuring and subsequent microsoldering for contacting, such that the nanostructures are free of resist as required for the scanning probe studies. It will be interesting to see in how far this novel approach will ease an optimization of graphene QDs, e.g., with respect to applications as qubits, where graphene has large fundamental advantages with respect to other materials as GaAs, but experimentally still lags significantly behind.

\section{acknowledgement}
We acknowledge the contributions from F. Libisch, R. Reiter, J. Burgd\"orfer, Y. Li, W. Zhang, R. Mazzarello, N. Attodiresei, S. Bl\"ugel,
P. Lazic, V. Caciuc, C. Busse, T. Michely, D. Subramaniam, V. Geringer, C. Pauly, A. Georgi, and T. Mashoff to the previous publications 
which are described in this review and the provision of figures by M. Fonin, P. Liljeroth, C. Busse, A. Varykhalov, F. Libisch, Y. Li, and R. Mazzarello.
We, moreover, thank A. Nent, L. Jung, B. Kaufmann, T. Kroesen, D. Zijlstra, and T. Hecking for their contributions to develop the anodic oxidation process used to prepare the graphene QDs. Financial support by the Graphene Flagship (Contract No. NECT-ICT-604391)
and by the German Science Foundation via LI 1050/2-2 is gratefully acknowledged.

%
%

\end{document}